\documentclass[twocolumn,nofootinbib,superscriptaddress]{revtex4-1}
\usepackage{graphicx,epsfig}
\usepackage{amsmath}
\usepackage{mathrsfs}
\usepackage{amsfonts}
\usepackage{braket}
\usepackage{xcolor}
\usepackage{hyperref}
\usepackage{fancyhdr}
\usepackage[normalem]{ulem}
\newcommand{\be}{\begin{eqnarray}}
	\newcommand{\ee}{\end{eqnarray}}
    
\newcommand{\p}{{\cal P}}
\newcommand{\As}{A_s}

\newcommand{\bb}{\Omega_{\mathrm{PBH}}}

\newcommand{\gammacr}{\gamma_{\mathrm {cr}}}
\newcommand{\Meq}{M_{\mathrm{eq}}}
\newcommand{\Mpl}{M_{\mathrm{Pl}}}
\newcommand{\keq}{k_{\mathrm{eq}}}

\begin{document}

	\title{Unexpected shape of the primordial black hole mass function}
	\author{Jacopo Fumagalli}
	\email{jfumagalli@fqa.ub.edu}
	\affiliation{Departement de F\'isica Qu\`antica i Astrofisica and Institut de Ci\`encies del Cosmos (ICC), Universitat de Barcelona, Mart\'i i Franqu\`es 1, 08028 Barcelona, Spain}
	\author{Jaume Garriga}
	\email{jaume.garriga@ub.edu}
	\affiliation{Departement de F\'isica Qu\`antica i Astrofisica and Institut de Ci\`encies del Cosmos (ICC), Universitat de Barcelona, Mart\'i i Franqu\`es 1, 08028 Barcelona, Spain}
	\author{Cristiano Germani}
	\email{germani@icc.ub.edu}
	\affiliation{Departement de F\'isica Qu\`antica i Astrofisica and Institut de Ci\`encies del Cosmos (ICC), Universitat de Barcelona, Mart\'i i Franqu\`es 1, 08028 Barcelona, Spain}
	\author{Ravi K. Sheth}
	\email{shethrk@physics.upenn.edu}
	\affiliation{Center for Particle Cosmology, University of Pennsylvania, 209 S. 33rd St., Philadelphia, PA 19130, USA}

	\begin{abstract}
	In a Universe with nearly-Gaussian initial curvature perturbations, 
 the abundance of primordial black holes can be derived from the curvature power spectrum. When the latter is enhanced within a narrow range around a characteristic scale, the resulting mass function has a single distinct peak, corresponding to Schwarzschild radii set by the horizon entry time of that scale. In contrast, we show (both numerically and by providing an analytic estimation) that a broad enhancement — such as a plateau bounded by infrared and ultraviolet scales — produces a bimodal mass function, with a primary peak close to the infrared scale. We find that the typical initial gravitational potential  (compaction function), conditioned on meeting the threshold for critical collapse, is generated by a thin spherical shell with infrared radius and a thickness comparable to the ultraviolet scale. This suggests a higher-than-expected abundance of PBH originating from Type II initial fluctuations. Our results significantly impact overproduction bounds on the amplitude of the power spectrum, and tighten the viable mass range for primordial black holes as dark matter.
 \end{abstract}
	
	\maketitle

\section{Introduction}

Little is known about the content of our Universe. While its evolution is well parameterized by assuming the presence of some form of dark matter (DM) and dark energy, the fundamental nature of these components remains elusive. A natural option for DM is that it only interacts gravitationally, and consists of primordial black holes (PBHs). Those may have been formed in the early universe, during the radiation dominated era, by gravitational collapse of cosmological perturbations of unusually large amplitude \cite{hawkingcarr}. In this scenario, the PBH abundance is tightly connected to the statistics of primordial perturbations, which for definiteness we shall assume to be of inflationary origin \cite{Ivanov:1994pa}.

PBHs are challenging to observe, particularly in the asteroid mass range where they could constitute the entirety of DM \cite{constr}. Current and future constraints combined with prior knowledge of the PBH mass function would therefore be invaluable in optimizing strategies for their detection. 
    
To date, the distribution of the cosmic microwave background fluctuations is compatible with nearly Gaussian statistics and an extremely small primordial power spectrum of curvature perturbations \cite{cmb}. This implies that, at cosmological scales, the probability of forming PBH via large random fluctuations is practically zero. Nevertheless, the power spectrum remains largely unconstrained at much smaller scales, potentially resulting in significant abundances of PBHs.

In this context, the standard expectation is that, for a nearly scale-invariant enhancement of the power spectrum between an infrared (IR) and an ultraviolet (UV) scale, the mass function of PBHs will be dominated by the lighter black holes, associated with the UV scale, which form earlier during radiation. 

Contrary to this expectation, in this Letter we show that, for nearly Gaussian initial curvature perturbations, heavier PBHs are {\em more} likely to form. This is due to a peculiar collective effect which occurs at smoothing scales comparable to the IR scale, leading to a dominant peak in the mass function. 
\section{Compaction function statistics}  
Consider a non-linear over-density in an asymptotically homogeneous and isotropic Friedmann-Lemaitre-Robertson-Walker (FLRW) background. In the case of spherical symmetry, we may define the Misner-Sharp excess mass \cite{ms} $\delta M(t,R)=4\pi\int_{0}^R \delta\rho(t,\tilde R) \tilde R^2 d\tilde R\Big|_t$, where $\delta\rho(t,R)\equiv\rho(t,R)-\rho_b(t)$ is the fluid over-density with respect to the background one ($\rho_b$). The integral is performed on a co-moving time slice orthogonal to the fluid lines and within a sphere delimited by the areal radius $R$. It is then useful to introduce the compaction function \cite{ss}
\be
	C(t,R)\equiv 2G\ \frac{\delta M(t,R)}{ R}\ .  
\ee   
The compaction function represents the inhomogeneous part of the local gravitational potential generated by the non-linear over-density. When $C$ is large enough, a gravitational collapse into a black hole is triggered by the formation of a trapped surface \cite{ss}. We are interested in a criterion for PBH formation in terms of the initial conditions for $C$. 

During inflation, perturbations are generated as quantum fluctuations of an inflaton field and become classical soon after their wavelength exits the (cosmological) horizon. If $L$ is the typical wavelength of the perturbation, and $H=\dot a/a$ is the Hubble rate of the background FLRW solution with scale factor $a(t)$, then the perturbed metric can be approximated in a gradient expansion on super-Hubble scales $L\gg H^{-1}$, as \cite{ss} 
 $ds^2\approx -dt^2 + e^{2\zeta(\vec{x})}a(t)^2d\vec{x}\cdot d\vec{x}$. 
Here $\zeta$ is the so-called co-moving curvature perturbation, encoding the spatial geometry of $t=\mathrm{const}$ hyper-surfaces. Note that $\zeta$ is independent of time on super-horizon scales.

The formation of a PBH is necessarily a rare event, requiring the local curvature to be well above its standard deviation. High peaks of a Gaussian random field are approximately spherically symmetric \cite{bbks}, and so we shall consider here a local super-horizon geometry of the form 
 $ds^2\approx -dt^2+e^{2\zeta(r)}a(t)^2\left(dr^2+r^2 d\Omega_2^2\right)$. 
In this case, the compaction function is time independent and can be written as \cite{japs}
\be
	C= g(r;\vec{x}_0) \Big(1-\frac{3}{8}g(r;\vec{x}_0)\Big)\label{cing}
\ee 
where we define $g=-(4/3)\,r\partial_r\zeta$. Thus, the superhorizon compaction function satisfies $C[g]\leq 2/3$, with $g=4/3$ saturating the upper bound. This critical value separates Type I ($g<4/3$) from Type II ($g>4/3$) perturbations.  In terms of the Fourier modes $\zeta_k$, and using the identity $r\partial_r\zeta=\frac{1}{4\pi r}\int d^3x\nabla^2\zeta\, \theta\left(r-\lvert\vec{x}-\vec{x}_0\rvert\right)$,
we have\footnote{Note that we are {\em not} assuming a window function in \eqref{g}. Using a different expression for $W$, as suggested by \cite{musso}, would not relate \eqref{g} correctly to \eqref{cing}.  Also, since our analysis uses quantities defined on super-horizon scales at a sufficiently early time, we have {\em not} introduced a transfer function as suggested in \cite{muscoriotto}.}  
\be
	g(r;\vec{x}_0)=\frac{4}{9}\int \frac{d^3 k}{(2\pi)^3} \, 
	e^{i \vec{k}\cdot \vec{x}_0}\, (kr)^2\, W(kr)\,\zeta_k\  , \label{g}
\ee
where
\be
 W(x)=3\frac{\sin(x)-x\cos(x)}{x^3}\ ,
 \label{Wth}
\ee
and with $\vec{x}_0$ we stressed that this is valid around a ``peak'' of $C$ centered at $\vec{x}_0$. Our assumption is that the Fourier modes $\zeta_k$ are independent and Gaussian-distributed. Because large fluctuations are increasingly less probable, Type I fluctuations are commonly expected to dominate the PBH abundance. Nevertheless, as we shall discuss, Type II fluctuations might not be such a far-off possibility. 

PBH formation occurs when the {\em initial} super-horizon $C(r_m;\vec{x}_0)$ at its maximum ($r=r_m$) exceeds a certain threshold $C_c$ \cite{musco}. Because we shall work on Type I fluctuations, this threshold can be immediately translated to the simpler Gaussian variable $g$. We can then ask what is the chance that  \cite{nnstat1,nnstat2}
\begin{itemize}
 \item [1)] $g(r;\vec{x}_0)$ has a peak as a function of position $\vec{x}_0$.
 \item [2)] $r_m$ is a local maximum of $g(r;\vec{x}_0)$, {\em i.e.} \be
 v(r_m)\equiv r_m\partial_r g(r_m;\vec{x}_0)=0\ee and \be w\equiv -r_m^2\partial_r^2 g(r_m;\vec{x}_0)
 \geq 0.\ee
 \item [3)] $g(r_m;\vec{x}_0)\geq g_c(w)$. 
\end{itemize}
The threshold $g_c$ depends significantly on the radial profile of the perturbation. It has been shown, however, that
for a wide class of of profiles, 
$g_c$ is approximately determined by the curvature $w$ at $r=r_m$. Ref. \cite{universal} (see also eqs. (22) and (58) of \cite{nnstat2}) gives an analytic approximation to $g_c(w)$ which matches the results of numerical simulations \cite{albert} to within a few percent.
For Type I black holes, it is a monotonic function with the limiting behavior 
\be \label{gc}
g_c(w\ll 1)\approx {1\over 2} \quad{\rm and}\quad  g_c(w\gg 1) \approx {4\over 3}-{32\over 9w}.
\ee
The abundance of PBHs will then be related to \cite{nnstat1,nnstat2} 
\be
	p(g,w,v=0) = p(v=0)\,p(g,w) 
	=\frac{p(g,w)}{\sqrt{2\pi \sigma_v^2}}, \label{general}
\ee
where the joint distribution of $w$ and $g$ is bivariate Gaussian. Setting $\sigma_{x}^2\equiv\langle x^2\rangle$, $\sigma_{xy}\equiv \langle xy\rangle$, 
$\gamma_{xy}\equiv\frac{\sigma_{xy}}{\sigma_{x}\sigma_{y}}$, 
and defining $\tilde{\sigma}^2_g = \sigma^2_g(1 - \gamma_{v g}^2), \tilde{\sigma}^2_w = \sigma^2_w(1 - \gamma_{v w}^2),$ and
\be 
\label{sigmastilde}
\tilde{\sigma}^2_{w g}  = \sigma^2_{w g} - \frac{\sigma_{w v}^2 \sigma_{v g}^2}{\sigma^2_v },
\ee 
Eq. \eqref{general} can be explicitly written as
\be
\label{pdfs}
p(g,w)&=& {\cal N}(g|0,\tilde{\sigma}^2_g)\,{\cal N}(w|\bar{w},\tilde{\sigma}_w^2(1-\tilde{\gamma}^2)) , 
\ee
where 
$\tilde{\gamma} = \frac{\tilde{\sigma}_{w g}^2}{\tilde{\sigma}_w \tilde{\sigma}_g}$ ,
$\bar w\equiv \tilde{\gamma}\frac{\tilde{\sigma}_w}{\tilde{\sigma}_g} g$ and ${\cal N}(x|\mu,\Sigma)$ is a normal distribution with mean $\mu$ and variance $\sqrt{\Sigma}$.
	
The correlators above are integrals over a power spectrum, defined by 
$\langle\zeta_k\zeta_{k'}\rangle = (2\pi)^3\delta(k+k')(2\pi^2/k^3) {\cal P}_\zeta(k)$.
 For example, $\langle g g \rangle\equiv \sigma^2_g = \sigma_0^2$, where 
\be\label{sigmag0}
 \sigma^2_j\equiv \frac{16}{81}\int \frac{dk}{k}\, (k r)^{4+2j}\, W^2(kr)\, \p_{\zeta}(k)\ .
\ee
Once $\p_\zeta$ is specified, the abundance of PBHs is (from now on we drop the sub-index ``$m$'' in $r_m$)
\be\label{abundance1}
\bb &\equiv  \int_{r_{\mathrm{min}
}}^{r_{\mathrm{eq}}} d\ln r \int_{0}^{\infty} dw  \int_{g_c(w)}^{4/3} dg  \, \mathcal{I}(g,w,r),
\ee
with 
\be\label{Integrand}
\mathcal{I}
\equiv  
\frac{w\,M(r,g,w)}{\rho_{\mathrm {eq}}a^3_{\mathrm{eq}}\,r^3}\,
   \frac{f\left(\chi/\sigma_\chi\right)}{(2\pi)^{3/2} (\sqrt{3}\,\sigma_1/\sigma_2)^3 }
\frac{p(g,w)}{\sqrt{2\pi\sigma_v^2}}\ ,
\ee
where 
$\rho_{\mathrm {eq}} =
3 \Mpl^2 H^2_{\mathrm {eq}}$ 
is the energy density at matter-radiation equality and $r_{\mathrm{min}}$ is the smallest scale associated with a black hole that has not yet evaporated. 
The factor $f\left(\chi/\sigma_\chi\right)$, is related to condition 1. It arises from a phase space integration over the traceless part of the Hessian, at the position $\vec x_0$ of the peak's center. The integral depends on the trace part, 
$\chi \equiv -r^2 \nabla^2_x g= 2g + w$, where the last equality follows from Eq.(\ref{g}) with $v(r)=0$. 

The explicit form of $f$ is given in Eq. (A15) of Ref. \cite{bbks}. The different $\sigma$'s can be expressed as combinations of $\sigma_j$ and their derivatives \cite{nnstat1}. 
The factor of $w$ arises from condition 2, as the Jacobian $dv/d\ln r$ at the extrema $v(r)=0$.  Condition 3 ensures that a PBH will form. 
Numerically, it has been found that the mass can be estimated as \cite{scaling}
\be\label{crisca}
M(r,g,w)\approx \mathcal{K}M_{H}(r)\bigl[C(g) - C(g_c(w)) \bigr]^{\gammacr}\ .
\ee
Here $M_H(r) =  4\pi\Mpl^2/ H(r)$  
is the Horizon mass, 
and the constant $\mathcal{K}\sim{\cal O}(1)$, which depends weakly on the profile shape (here parametrized by $w$), captures our ignorance of the local Hubble scale. To illustrate our results, we set $\mathcal{K}\simeq 6$ and  $\gamma_{\rm cr}\simeq 0.36$ \cite{universal}. 

\section{Bimodality from the Broad power spectrum}	
The two typical limiting cases studied in the literature regarding the spectrum of enhanced primordial perturbations, correspond to very narrow or very broad $\p_\zeta$. While a narrow spectrum has the virtue of mathematical simplicity, a broad spectrum is a more natural outcome. If we set  
\be\label{tophat}
\mathcal{P}_{\zeta}(k) 
= \As\, \theta(k-k_{\rm IR})\,\theta(k_{\rm UV} - k),
 \label{Pzeta}
\ee
then `broad' means $\alpha\equiv k_{\rm UV}/k_{\rm IR}\gg 1$\footnote{We use the power spectrum shape given in Eq. \eqref{tophat} to maintain   analytical control. However, we have also numerically verified the robustness of our findings for $\mathcal{P}_\zeta$ without sharp cutoffs.}. More specifically, we will see that broad and narrow spectra produce large differences whenever $\alpha\gtrsim 25$ for relevant values of the power spectrum amplitude. In terms of e-folds ($a=e^{N}$), this implies $N \gtrsim 3$. Thus, a wide class of inflationary models related to PBH formation falls into this category.
\begin{figure}
\centerline{\includegraphics[width=0.44\textwidth,angle=0]{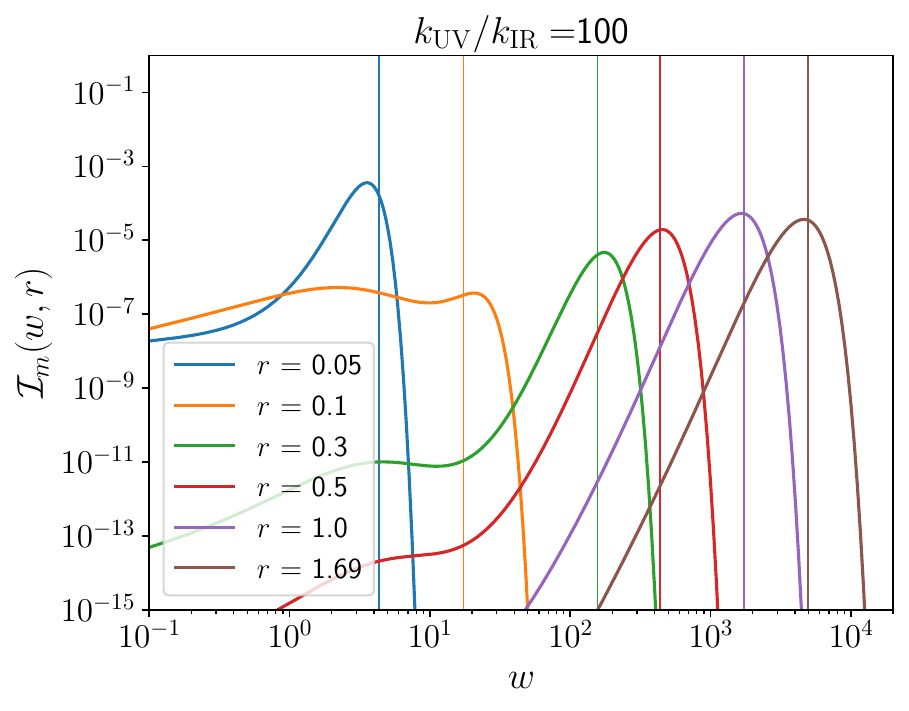}}
\caption{PDF marginalized over $g$ as a function of $w$ for different $r$ (in units $1/k_{\mathrm{IR}}=1$) for $\alpha = 100$ and $\As = 10^{-2}$. Vertical lines show our analytic estimate of the maximal $w$, Eq.\eqref{wmax}. The PDF is normalized by the factor $k_{\rm IR}/\keq$.} 
\label{pdf_marginalized_w}
\end{figure}
To understand PBH abundances for a broad spectrum, it is useful to first reconsider the narrow case ($\alpha\to 1$) with a peak at the UV scale. There, the abundance is sharply peaked around masses $\sim M_H(r_c)$ where $r_c\simeq 2.74/k_{\rm UV}$ is the first zero of $\sigma_v$ \cite{nnstat1} (c.f. Appendix for a thorough discussion).  As $\alpha$ increases, the ``UV" peak in the mass function persists, though it shifts to larger scales before saturating at $r\simeq 4/k_{\rm UV}$ for $\alpha\gtrsim 1.5$ ($N\gtrsim 0.4$).  Moreover, as we discuss below, a second peak related to the IR scale is generated, making the predicted distribution bimodal. 
	
Setting units with $r_{\mathrm{IR}}=1/k_{\mathrm{IR}}=1$, we now consider $r\alpha \gg 1$. In this limit, one finds that all correlations with $v$ become subdominant and $\tilde{\gamma}\simeq \gamma_{w g}\equiv \gamma \simeq 1/\sqrt{\ln \alpha}\ll 1$. The latter inequality defines what we mean by a ``broad'' spectrum. Additionally, in this limit, the variances in Eq. \eqref{sigmastilde} become $\tilde{\sigma}_{X}\simeq \sigma_{X}$ for all variables.

Importantly, although $\gamma$  is small in this regime, 
$\bar w/  \sigma_w\,=\gamma\, (g/\sigma_g)\propto 1/(\sqrt{\As}\ln \alpha)$ is sizable due to the smallness of $\sigma_g \propto \sqrt{A_s\ln\alpha}$ and $\sigma^2_w \simeq (2/9) \As  (r\alpha)^4$. Thus, the PDF of $w$ for $r\alpha\gg 1$ develops a maximum for \textit{large} values of $w$. On the other hand, as the threshold grows with $w$, the statistics of $g$ would favor smaller values of $w$. Nevertheless, as long as $\sigma_g$ is not too small, the peak theory function $f$ will help suppress small $w$ configurations (as we shall discuss soon), so that profiles with $w\sim \bar{w}$ will be most probable (see Fig. \ref{pdf_marginalized_w}). 
   
One can estimate when that happens by comparing the integrand Eq.\eqref{Integrand} in two regimes: $r\sim \alpha^{-1}$ ($w\ll 1$) and $r\alpha\gg 1$ ($w\gg 1$). Neglecting the critical scaling for simplicity, using $f\approx g^8/\sigma_w^8\,(w\ll1)$ and $f\approx w^3/\sigma_w^3\,(w\gg1)$, the condition  $\mathcal{I}(w=0,g=g_c(0)=1/2,r=1)\lesssim \mathcal{I}(w=\bar{w},g=g_c(\infty)=4/3,r=1)$ leads to 
\be\label{condition}
A_s (\ln \alpha)^2\gtrsim 0.05.
\ee
Whenever the previous condition is satisfied, profiles with $w \gg 1 $ dominate the statistics for $r\alpha \gg 1$.
For the regime $r\alpha \gg 1$ and for amplitudes high enough such that $w\gg 1$ (where $g_c$ is close to the boundary between Type I and II black holes), we can estimate Eq. \eqref{abundance1} as follows.  
Using the expansion of the critical threshold in Eq. \eqref{gc}, and Taylor expanding the integral in $g$, yields 
\be\label{intw}
	\mathcal{I}_m(w,r) \propto w^{\lambda} \exp\left(-\frac{(w - \gamma \sigma_w\nu_c )^2}{2\sigma_w^2}\right) e^{-\frac{\nu_c^2}{2}},
\ee
where we have defined 
\be
\nu_c(r) \equiv\frac{ 4/3}{\sigma_g(r)}
\ee  
to compare with the often-quoted Press-Schechter case \cite{ps}. Because $w$ is large, we can approximate $f(x)\simeq x^3$ finding $\lambda \simeq 2.28$.  Had we omitted the peak theory weightings (or the Jacobian proportional to $w$), we would have obtained $\lambda = -1$ and would have concluded that the PDF is dominated by broad $C(r)$ profiles (see e.g. \cite{broader}).

Eq.\eqref{intw} is maximized at
\be\label{wmax}
\frac{w_{\rm max}}{\sigma_w}=\frac{\gamma\nu_c+\sqrt{\gamma^2\nu_c^2+4\lambda}}{2}  ,
\ee 
where $\gamma\nu_c\sigma_w \simeq (2/3)(r\alpha)^2/ \ln \alpha$.  I.e., $w_{\rm max}\gg 1$ when $r\alpha \gg 1$.  
Fig. \ref{pdf_marginalized_w} shows that Eq. \eqref{wmax} with $\lambda = 2.28$ provides a good estimate of the $w$ at which the marginalized PDF is maximized.

Next, using the saddle point approximation centered on $w_{\rm max}$, gives 
\be\label{domegadlogr}
 \frac{d \Omega_{\mathrm{PBH}}}{d \ln r} = 
  B \, \left(\frac{r}{r_{\mathrm{UV}}}\right)^{1-4\gammacr}\, 
  \nu_c^{4-4\gammacr}\wp_L(\nu_c),
\ee
where $\wp_L(\nu_c)=\frac{1}{2}\nu_c^3 e^{-\frac{\nu_c^2}{2}}$ 
is the `high peaks' scaling of the linear smoothed over-density \cite{bbks, germanimusco}
and 
\be
B = \frac{r_{\mathrm{eq} }}{r_{\mathrm{UV}}}\As^{1-2\gammacr}\frac{\mathcal{K} \,2^{7\gammacr+1}}{(2\pi)^{3/2}3^{\gammacr+5/2}} \ ,
\ee
with $r_{\rm eq}\equiv 1/(a(t_{\rm eq})H_{\rm eq})$.
Importantly, $\nu_c$ is $r$-dependent via $\sigma_g^{-1}$ and this matters for the maximization of $d\bb/d\ln r$. Thus, expanding at leading (polynomial) order in $\alpha$ we have
\begin{equation}\label{sigmag}
\sigma_g^2\simeq\frac{8}{9}A_s\left(\ln\alpha+{\rm CosIntegral}(2r)+\frac{\sin^2 r-r\sin (2r)}{r^2}\right).
\end{equation}
This function has a maximum -- independent of $\alpha$ -- at $r\simeq 1.69$ (in units of $r_{\mathrm {IR}}= 1/k_{\mathrm{IR}}$), which gives rise to a large contribution around IR scales. 

\section{Mass function}
Let us define what is customarily -- but not exclusively -- called the mass function $f_{\mathrm {PBH}}(M)$: the fraction of energy density contained in primordial black holes per logarithmic mass interval, such that $\Omega_{\mathrm{PBH}} = \int d\log M f_{\mathrm {PBH}}(M)$. 

By inverting Eq. 
\eqref{crisca}, we may express $g$ 
as a function of $(M,w,M_H)$. Denoting this function by $g_{\bullet}$, we get \cite{nnstat2}:
\begin{align}
 f_{\mathrm{PBH}}(M) &=
  \left(\frac{M}{\mathcal{K}M_{\rm eq}}\right)^{\frac{1}{\gammacr} +1}
 \int \frac{dM_H}{M_H} 
 \left(\frac{M_{\rm eq}}{M_H}\right)^{\frac{1}{\gammacr} +1}
 \nonumber\\
 &\quad \times \frac{2\pi\mathcal{K}}{3}\int_0^{\infty} dw \frac{w}{\gammacr (1-\frac{3}{4}g_\bullet)}\left(\frac{M_ {\rm{eq}}}{M_H}\right)^{1/2}
 \nonumber\\ 
 &\quad \times p(g_{\bullet},w,v=0)\frac{f\left(\frac{2g_{\bullet}+w}{\sigma_\chi}\right)}{(2\pi)^{3/2} (\sqrt{3}\sigma_1/\sigma_2)^3 }\ ,\label{massfu}
 \end{align}
where $M_{\rm eq}$ is the horizon mass at matter-radiation equality.
\begin{figure}
\includegraphics[width=0.48\textwidth,angle=0]{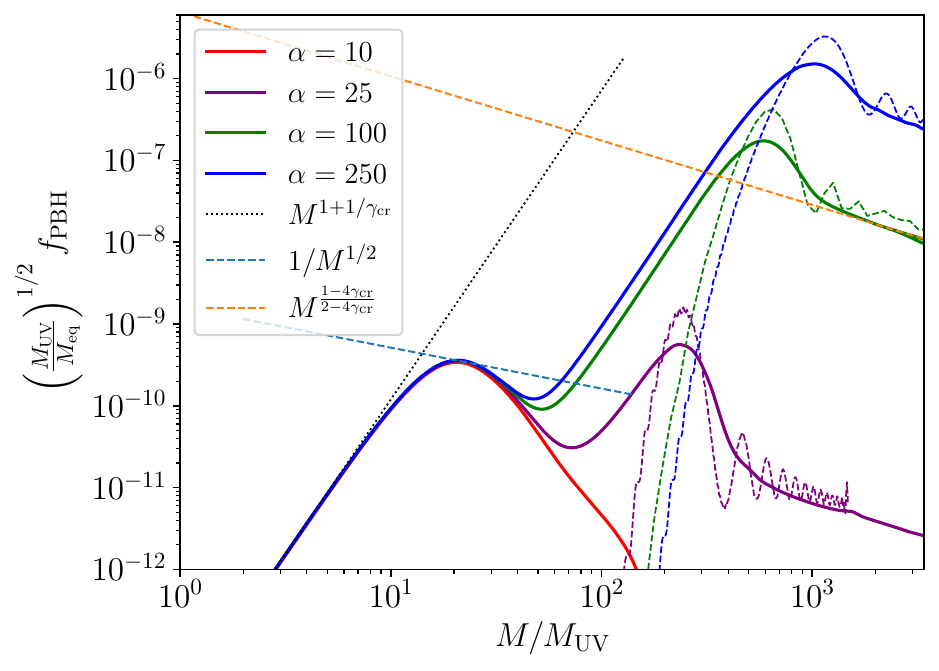}
\caption{Bimodal PBH mass functions for flat power spectra (Eq. \eqref{tophat}) and a range of $\alpha \equiv k_{\mathrm{UV}}/k_{\mathrm{IR}}$ as labeled ($\As = 10^{-2}$ in all cases). Solid curves show the exact result (Eq. \ref{massfu}), while dashed curves show the approximation (Eq. \ref{approxfPBH}) valid for IR scales ($r/r_{\mathrm{UV}} \gg 1$), both normalized by $(\Meq / M_{UV})^{1/2}$, where $M_{\mathrm{UV}} \equiv M_H(k_{\mathrm{UV}})$.
As $\alpha$ increases, a second feature in the mass function arises in the IR: the two peaks become more widely separated and the one at higher masses dominates.
The IR peak scales generically as $M_{\mathrm{heavy}}/ M_{\mathrm{UV}} \propto (k_{\mathrm{UV}}/k_{\mathrm{IR}})^{2-4\gammacr}$. 
The standard expectation for a broad spectrum is also displayed. There, the UV peak was expected to grow with the critical scaling (dotted black line) along with a subsequent power-law falloff $M^{-1/2}$ (dashed blue line) due to red-shift dilution, see e.g. \cite{Byrnes:2018clq,riottobroad}.}
\label{mass}
\end{figure}

As shown in in Figure \ref{mass}, the mass function develops a peculiar bimodal shape as $\alpha$ increases, with the following features:
\begin{itemize}
\item\textbf {First Peak:} In the UV, where $r\sim 1/k_{\rm UV}$, $f_{\mathrm{PBH}}$ is dominated by the critical scaling from the first line of Eq. \eqref{massfu} (dotted lines in Fig.~\ref{mass}). This leads to a first peak, corresponding to the UV scale of the power spectrum related to the first maximum of $\tilde\sigma_w(r)$. We find that generically this is for $r\approx 4/k_{\mathrm{UV}}$.

\item\textbf{Second Peak:} 
After the first peak, we have an initial decay of the mass function that is steeper than the usual power-law falloff dilution ($\propto M^{-1/2}$).   
The intermediate drop between the two peaks, which arises from the higher threshold associated with those profiles, is eventually compensated by the growth of $\tilde{\sigma}_g$ together with the decay of $\gamma$'s, until reaching a maximum at $r=1.69 /k_{\mathrm{IR}}$ (c.f. Eq. \ref{sigmag}). 
There, profiles with large $w$ dominate the statistics.
\end{itemize}
It is useful to estimate the IR (large mass) part of $f_{\rm PBH}$. Inverting $r$ from the critical scaling Eq. \eqref{crisca} and applying the results from the previous section, i.e. $ w \approx \bar{w} $ and $ g \approx 4/3 $, one obtains $ r \approx r(M, \bar{w}, g = 4/3) $. In this way $ r $ can be approximately written as a function of $ M $ and (using $k_{\rm eq}\equiv 1/r_{\rm eq}$)
\be \nonumber
 f_{\mathrm{PBH}}(M)
 &\simeq& \tilde{B}\, 
 \left(\frac{M_{\mathrm{UV}}}{M}\right)^{\frac{4\gammacr-1}{2-4\gammacr}}
 \nu_c^{5-4\gammacr}\wp_L(\nu_c) \\[1mm] &\propto& M^{-0.79}
 \nu_c^{3.56}\wp_L(\nu_c),
 \label{approxfPBH}
\ee
where $M_{\mathrm{UV}} \equiv M_H(k_{\mathrm{UV}})$, we introduce the coefficient
\be\label{tildeB}
\tilde{B} = B\frac{1}{(2-4\gammacr)}\left(\mathcal{K} \frac{2^{5\gammacr}}{3^{\gammacr}}\left(\ln \alpha\right)^{2\gammacr}\right)^{\frac{4\gammacr-1}{2-4\gammacr}},
\ee
and we set  $\gammacr\simeq 0.36$ in the last line.
Note that the largest dependence of the peak amplitude of $f_{\rm PBH}$ resides in $\nu_c\propto\sigma_g^{-1}\propto (\As\ln\alpha)^{-1/2} $ appearing in $\wp_L(\nu_c)$.

Eq. \eqref{approxfPBH} captures the behavior of $f_{\mathrm{PBH}}$ in the infrared, as shown in Figure \ref{mass}. Larger values of $\alpha$ correspond to a greater abundance of heavier PBHs. In the IR, $\nu_c$ settles to a constant, and $f_{\mathrm{PBH}}$ follows a power-law falloff, as described by Eq. \eqref{approxfPBH} (see Figure \ref{mass}).

The PBH mass associated with this second maximum is smaller than the horizon mass related to the IR scale. This is because the departure from criticality (Eq.\ref{crisca}) decreases as $w$ increases. It is straightforward to estimate that $M_{\mathrm{heavy}}/ M_{\mathrm{UV}}= \mathcal{K}\, (k_{\mathrm{UV}}/k_{\mathrm{IR}})^2\, (1.69)^2\, (C(4/3)-C_c(w_{\mathrm{max}}))^{\gammacr}\propto\alpha^2\, w_{\rm max}^{-2 \gammacr} \propto \alpha^{2-4\gammacr}$.

Note that the double-peak structure described above holds as long as the power spectrum amplitude is sufficiently large, as quantified by Eq. \eqref{condition}. For small values of $A_s$ (or equivalently $\alpha$), we have numerically checked that the bimodal structure merges into a single peak. This is because the statistics gets dominated by the small $w$ profiles. There, $f \propto \sigma_w^{-8} \propto 1 / (r\alpha)^{16}$, which strongly disfavors infrared profiles. In addition, the approximation of spherical symmetry used so far would hardly hold in this regime. In any case, parameter choices which result in unimodal mass functions turn out to be irrelevant in terms of abundance. For instance, for $A_s \lesssim 0.003$ and $\alpha = 100$, even assuming $M_H(k_{\rm UV}) \sim 10^{17} \, \rm g$, which is the smallest possible PBH mass compatible with the Hawking evaporation bounds, yields  
$f_{\mathrm{PBH}} \lesssim 10^{-7}$.

\section{Outlook}
When the power spectrum of primordial perturbations is enhanced across a broad range of scales, a significant contribution to the PBH mass function emerges from heavy black holes associated with the infrared scale.  This is quite natural in our statistics as the required two main conditions: 1) the existence of a peak in $\vec{x}=\vec{x}_0$ and 2) the maximum of the compaction function at $r=r_m$, become more uncorrelated for larger distances $r_m$.

In this Letter, we have highlighted this phenomenon, which arises from over-densities in the form of thin (UV-scale) spherical shells, characteristic of the typical compaction function profile. This additional dominant contribution has the potential to reshape constraints from PBH overproduction from 
gravitational wave signals. Also, even for moderately broad power spectra, a bimodal distribution arises which effectively narrows the viable range for PBH as the dominant form of DM, particularly in the asteroid-mass window. The reason is that, since the subdominant  UV population has to avoid the evaporation bounds, the dominant IR population must be shifted to values which are significantly higher than the existing lower bound in this window. A more thorough quantitative analysis of these effects is left for future study.

Our findings also reveal that, in the presence of a broad spectrum, the statistics of initial perturbations that will collapse into PBH are pushed to the boundary between Type I and Type II. This calls for a deeper understanding of PBH formation through Type II fluctuations, which are often ignored since they require higher (and hence exponentially suppressed) values of the over-density compared to the more standard Type I fluctuations--see \cite{type2} for recent numerical studies. 
Additionally, dedicated simulations of gravitational collapse for configurations with large values of $w$, involving in turn large separation of scales, are needed to confirm the extrapolation \cite{musco,albert} of a threshold saturation in the Type I case.

\begin{acknowledgments}
The authors are grateful to Albert Escriv\`a for guidance on the latest numerical results in PBH formation.  
The research of JF, JG and CG is supported by the
grant PID2022-136224NB-C22, funded by MCIN\allowbreak/\allowbreak AEI\allowbreak/10.13039\allowbreak/501100011033\allowbreak/\allowbreak FEDER,
UE, and by the grant\allowbreak/ 2021-SGR00872.  RKS is grateful to the ICTP for hospitality in 2024.
\end{acknowledgments}

\appendix

\section*{Supplemental Material}
	
\section{Sharply peaked power spectrum}

In this Appendix, we discuss the PBH abundance in the  limit where the enhancement in the power spectrum is very narrow. We derive an analytic expression for the mass function which matches the numerical results obtained from Eqs. (\ref{abundance1}) and (\ref{Integrand}) in the limit $(\alpha-1) \ll 1$. More specifically, we consider a power spectrum of the form 
 \begin{equation}
 {\cal P}_\zeta(k) =\mathscr{A} k_0 \delta(k-k_0),
  \label{eq:sharpPk}
 \end{equation}
 which can be obtained from Eq. (\ref{tophat}) in the limit $\alpha\to 1$ and $A_s\to \infty$, while keeping $\mathscr{A}\equiv  A_s\ln\alpha $ finite. 
 
This limit can be done directly on Eqs. (\ref{abundance1}) and (\ref{Integrand}), but the expansions become somewhat cumbersome. Here, we present an equivalent alternative route which is more economical for the case at hand. 

First, note that the correlators of interest in the main text become 
\begin{align}
 \sigma_j^2(r) &= \frac{16}{81}\mathscr{A}\,\kappa^4\,W^2(\kappa)\,\kappa^{2j}, \quad
 \sigma_w^2(r) = (\kappa^2 - 2)^2\,\sigma_0^2 , \nonumber\\ \label{eq:sharpSigmas}
 \sigma_v^2(r) &= \frac{16}{81}\mathscr{A}\kappa^4\,\Big(3j_0(\kappa) - W(\kappa)\Big)^2, \qquad{\rm and}\\  
 \langle gw\rangle &= \sigma_0\sigma_w, \quad 
 \langle wv\rangle = \sigma_w\sigma_v \quad {\rm and}\quad
 \langle gv\rangle = \sigma_0\sigma_v. \nonumber
\end{align}
(Here $\kappa\equiv k_0r$ and $W(\kappa)$ is given by Eq.~\ref{Wth}.)
Therefore, 
\begin{equation}\label{proportionality}
 g/\sigma_g = w/\sigma_w = v/\sigma_v, 
\end{equation}
for generic values of $r$.

Next, we backtrack to the expression for the number density of peaks per co-moving volume of the random variable $g_r(\vec x_0)\equiv g(r;\vec x_0)$, which are characterized by
 $\chi_r = -r^2 \nabla^2_{\vec x_0} g_r(\vec x_0)$, $v_r =r \partial_r g_r$, and $w_r=-r dv_r/dr$ \cite{bbks,nnstat2}:
\begin{equation}
dn = {f(\chi_r/\sigma_\chi)\over (2\pi)^{3/2} r_*^3}\,   p(g_r,w_r,v_r=0)\, 
    dv_r dw_r dg_r,\label{numberdens}
\end{equation}
where $r_* = r\, (\sqrt{3}\,\sigma_1/\sigma_2)$.
We need not integrate over $\chi_r$, since it follows from Eq. (\ref{g}) that it is maximally correlated with $2g_r+w_r$ \cite{nnstat1}, 
\begin{equation}
p(\chi_r |g_r,w_r) = \delta(\chi_r -2g_r-w_r).\label{chidelt}
\end{equation} 
We are interested in the value of $r$ corresponding to an extremum of $g_r$, i.e. $v_r =0$. Therefore, we may trade the integration over $v_r$ for an integration over $r$, using $dv_r/dr = -(w_r/r)$. 

In the main text, we express 
$p(g_r,w_r,v_r=0) = p(g_r,w_r|v_r=0)p(v_r=0)$, which leads to our expression for $\Omega_{\rm PBH}$, Eq. (\ref{abundance1}).   
In the present case, where the power spectrum is monochromatic, Eq. (\ref{g}) also implies that all realizations of $g$ have the same $r$ dependence. Therefore, the radial derivatives $v_r$, $w_r$ are maximally correlated with $g_r$,
\be
\quad p(w_r|g_r) &=& \delta(w_r - (\sigma_w/\sigma_g) g_r),\cr 
p(v_r|g_r) &=& \delta(v_r - (\sigma_v/\sigma_g) g_r), \label{tonadeltas}
\ee
which is consistent with Eq.\eqref{proportionality}.
%\begin{equation}\label{proportionality}
%g/\sigma_g=w/\sigma_w=v/\sigma_v, 
%\end{equation}
%for generic values of $r$.
For $g_r\neq 0$, Eq.\eqref{proportionality} implies that $v_r=0$ if and only if $\sigma_v(r)=0$, which occurs when $3j_0(\kappa)=W(\kappa)$ for all realizations of the random field with spherical symmetry (Eq.~\ref{eq:sharpSigmas}). This has multiple zeros, of which the first is at $r=r_c\simeq 2.74/k_{0}$. Note also that, from the last equation in (\ref{tonadeltas}), we have
\begin{equation}\label{deltarc}
 p(v_r=0|g_r\neq 0) dv = \delta(r-r_c)dr,
\end{equation}
where we have used $\sigma_w=|r\partial_r \sigma_v|$ and Eq. (\ref{proportionality}). Hence, the $v$ integration reduces to substituting $r=r_c$ everywhere in the integrand.  Since $r$ is a like a time variable, this indicates that, for a monochromatic $P_\zeta$, all PBHs form at the same time.  

However, they form with a range of masses.  To see this explicitly, use (\ref{deltarc}) and (\ref{tonadeltas}) in (\ref{numberdens}) to obtain
\begin{equation}
d \Omega_{\rm PBH} = {M\over \rho_{\rm eq} a_{\rm eq}^3 } dn 
= {M\over \rho_{\rm eq} a_{\rm eq}^3}\, {f(g/\sigma_g)\over (2\pi)^{3/2} r_*^3} \,p(g)\, dg,
\end{equation}
where we have dropped the subindex $r$. In the previous expression $p(g)$ is a centrally peaked Gaussian.  By inverting (\ref{crisca}), we have $g=g_\bullet(M)$, where
\begin{equation}
 g_\bullet(M) = {4\over 3}\left(1-\sqrt{1-{3\over 2}C_c-{3\over 2} \left({M\over \mathcal{K} M_H}\right)^{1\over\gamma_{cr}}}\right). 
\end{equation}
Here, $C_c=C(g_c)$. Since $g=(\sigma_g/\sigma_w)\ w$, the Type I PBHs condition $g_c(w)<g <4/3$ leads to $g_c(w)\simeq[0.91,0.98]$. Because of the exponential suppression in $p(g)$ it will be sufficient to consider the lower limit $g_c\approx 0.91$. Since $(\sigma_w/\sigma_g) = \kappa^2 - 2$ (Eq.~\ref{eq:sharpSigmas}), at $r_c$ it is $\approx 2.74^2-2 = 5.52$, so the value $g_c\approx 0.91$ implies $w\approx 5.1$.  I.e., as in the case of the broad spectrum, the most relevant compaction function for PBH of Type I is {\em not} broad, in contrast to some recent claims \cite{broader}. Rather, it is more like a thin spherical shell, in good agreement with recent numerical work on configurations associated with stochastic fluctuations in ultra-slow-roll inflation \cite{rrtNarrow}.

Finally, by expressing $dg$ in terms of $d\ln M$ we find the mass function (as defined in the main text) for a sharply peaked power spectrum:
\begin{equation}\label{approsharp}
f_{\rm PBH} = D\  
{f\left({g_\bullet\over \sigma_g}\right) e^{-{g_\bullet^2\over 2\sigma_g^2}} \over  \gamma_{cr}[1-(3/4) g_\bullet] \sigma_g} \left({M\over {\cal K} M_H}\right)^{1+{1\over \gamma_{\rm cr}}},
\end{equation}
where
\begin{equation}
 D = \frac{\mathcal{K}M_H}{(2\pi)^{2}  \rho_{\rm eq} a_{\rm eq}^3r_*^3}.
\end{equation}
\vspace{-0.5em}

\begin{figure}[htbp]
\begin{center}
\includegraphics[width=0.46\textwidth]{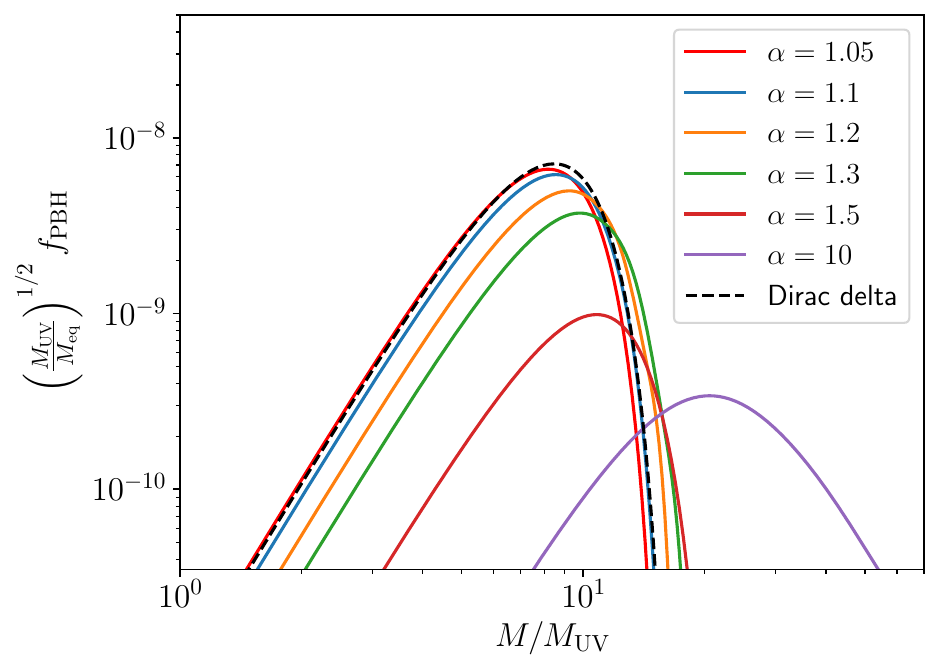}
\caption{Mass function for narrow power spectra with a fixed logarithmic area $\mathscr{A} = 10^{-2}$ and varying $\alpha = k_{\mathrm{UV}} / k_{\mathrm{IR}}$, as indicated in the legend. The exception is the not-narrow case of $\alpha = 10$, with height $A_s = 10^{-2}$ (and so $\mathscr{A}\simeq 0.023$), whose parameters are taken from Figure \ref{mass} and shown for comparison. The dashed line represents the analytical approximation Eq. \eqref{approsharp}, for a Dirac delta power spectrum (Eq.~\ref{eq:sharpPk}) centered at $k_{\rm 0} = k_{\mathrm{UV}}$.}
\label{app}
\end{center}
\end{figure}

\noindent The value of the coefficient can then be expressed as
\begin{equation}\nonumber
D= \frac{(4\pi/3)\,\mathcal{K}}{(2\pi)^2(\sqrt{3}\,\sigma_1/\sigma_2)^3}
\sqrt{\frac{\Meq}{M_H(r_c)}}
%\left(\frac{\Meq}{M_H(r_c)}\right)^{\frac{1}{2}}
\simeq 2.54 \left(\frac{\Meq}{M_H(r_c)}\right)^{\frac{1}{2}},
\end{equation}
where the final expression used $\sigma_2/\sigma_1 = \kappa = 2.74$ (from Eq.~\ref{eq:sharpSigmas}) and ${\cal K}=6$.  To make contact with the main text, we may take $k_0 =k_{\rm UV}$, and note (again from Eq.~\ref{eq:sharpSigmas}) that $\sigma_g^2 \approx 2.01\mathscr{A}$ when $\kappa=2.74$.
% k0 r* = k0 r sqrt(3)/kappa -> sqrt(3) 
Fig. \ref{app} shows good agreement between the approximation \eqref{approsharp} and the full numerical calculation.  Note that the UV peak slowly drifts to the IR as $\alpha$ is increased, before reaching its asymptotic value, where it saturates for all $\alpha\gtrsim 3$. In terms of e-folds, we have that for $N\gtrsim 1$ the approximation of a narrow spectrum is no longer valid.

\end{document}